\documentclass[notitlepage,letterpaper,showpacs,preprintnumbers,amsmath,nofootinbib,amssymb,twocolumn, superscriptaddress, hyperref]{revtex4-1}

\pdfoutput=1
\usepackage{amssymb,amsmath,latexsym,mathrsfs}
\usepackage{graphicx} 
\usepackage{bm}
\usepackage[normalem]{ulem}
\usepackage[usenames,dvipsnames]{color}
\usepackage{hyperref}
\hypersetup{
    colorlinks=true,       
    linkcolor=NavyBlue,          
    citecolor=Magenta}   
\usepackage{soul,xcolor}

\newcommand\ee{\end{equation}}
\newcommand\be{\begin{equation}}
\newcommand\eea{\end{eqnarray}}
\newcommand\bea{\begin{eqnarray}}

\newcommand\vev[1]{{\langle {#1} \rangle}}

\renewcommand{\eprint}[1]{[\href{http://arxiv.org/abs/#1}{\tt#1} [astro-ph.CO]]} 
\newcommand{\eeeprint}[1]{[\href{http://arxiv.org/abs/#1}{\tt#1}] }
\newcommand{\eeprint}[1]{[\href{http://arxiv.org/abs/#1}{\tt#1} [astro-ph]} 
\newcommand{\ep}[2]{[\href{http://arxiv.org/abs/#1}{\tt#1} {#2} }

\newcommand\ie{{\it i.e.}~}
\newcommand\eg{{\it e.g.}~}

\renewcommand\d[1]{\:\textrm{d}#1}

\renewcommand\({\left(}
\renewcommand\){\right)}
\renewcommand\[{\left[}
\renewcommand\]{\right]}

\begin{document}
\preprint{IFIC/15-64}

\setstcolor{red}

\title{The present and future of the most favoured inflationary models after \emph{Planck} 2015}

\author{Miguel Escudero} 
\affiliation{
Instituto de F\'isica Corpuscular (IFIC), CSIC-Universitat de Valencia,\\ 
Apartado de Correos 22085,  E-46071, Spain}

\author{H\'ector Ram\'irez}
\affiliation{ Instituto de F\'isica Corpuscular (IFIC), CSIC-Universitat de Valencia,\\
Apartado de Correos 22085,  E-46071, Spain}

\author{Lotfi Boubekeur}
\affiliation{ Instituto de F\'isica Corpuscular (IFIC), CSIC-Universitat de Valencia,\\
Apartado de Correos 22085,  E-46071, Spain}
\affiliation{Universidad San Francisco de Quito USFQ, Colegio de Ciencias e Ingenier\'ias El Polit\'ecnico, campus Cumbay\'a, calle Diego de Robles y V\'ia Interoc\'eanica, Quito EC170157, Ecuador.}

\author{Elena Giusarma}
\affiliation{Physics Department and INFN, Universit\`a di Roma ``La Sapienza'', Ple Aldo Moro 2, 00185, Rome, Italy}

\author{Olga Mena} 
\affiliation{
Instituto de F\'isica Corpuscular (IFIC), CSIC-Universitat de Valencia,\\ 
Apartado de Correos 22085,  E-46071, Spain}

\begin{abstract}
{The value of the tensor-to-scalar ratio $r$ in the region allowed by the latest {\sl Planck 2015} measurements can be associated to a large variety of inflationary models. We discuss here the potential of future Cosmic Microwave Background cosmological observations in disentangling among the possible theoretical scenarios allowed by our analyses of current \emph{Planck} temperature and polarization data. Rather than focusing only on $r$, we focus as well on the running of the primordial power spectrum, $\alpha_s$ and the running thereof, $\beta_s$. If future cosmological measurements, as those from the COrE mission, confirm the current best-fit value for $\beta_s \gtrsim 10^{-2}$ as the preferred one, it will be possible to rule-out the most favoured inflationary models.}

\end{abstract}

\maketitle

\twocolumngrid

\section{Motivations} 
The smoking-gun of inflation~\cite{Guth:1980zm,Linde:1981mu,Albrecht:1982wi} is the detection of a stochastic background of gravitational waves. Such primordial signature is characterized by its amplitude, parametrized via the  tensor-to-scalar ratio $r$. Recent analyses from {\sl Planck} 2015~\cite{planck} have presented the tightest bounds to date on $r$ using temperature and polarization measurements. Albeit current {\sl Planck} constraints are perfectly compatible with a vanishing tensor-to-scalar ratio, yet there is still enough room for other theoretical possibilities besides the Starobinsky $R^2$-gravity scenario, which emerges as the best-fit model. Looking forward to the next generation of CMB observations, and depending on the value of $r$ that Nature has chosen, one can envision two distinct possibilities: \emph{(a)} either $r$  turns out to be way too small to be measured by the next generation of CMB observations, or \emph{(b)} the value of $r$ is large enough to be detected. However, in this latter case, the measured tensor-to-scalar ratio will typically correspond to several inflationary models. Given that measuring $r$ (if $r <few \times 10^{-4}$) might be extremely difficult~\cite{lowestlensing,lowestforeground}, and disentangling between the various models that lie in the same regions in the canonical $(n_s, r)$ plane might not be straightforward either, we explore here the possibility of extending the analysis to other (complementary) inflationary observables.

For the scalar power spectrum of the primordial perturbations, we consider,  as additional observables, the running $\alpha_s$ and the running of the running $\beta_s$. For the primordial tensor power spectrum, we consider its running $n_t$. The aim of this paper is to assess the potential of future CMB observations in falsifying inflation (or unraveling the fundamental model among the most favoured candidates after  {\sl Planck 2015}  data) by looking to these three additional observables. For illustration, we will consider some well-motivated models that are compatible with current data.
The structure of the paper is as follows. Section~\ref{sec:basic} deals with the basic definitions of the different cosmological observables and their current constraints. Section \ref{sec:models} describes the theoretical predictions from the most favoured inflationary scenarios after {\sl Planck} 2015 CMB temperature and polarization measurements. In Section~\ref{sec:current} we perform Markov Chain Monte Carlo (MCMC) analyses of the \emph{Planck} 2015 data release. Our Fisher matrix forecasts in Section~\ref{sec:future} show that, if the future preferred value of $\beta_s$ is close to the current best-fit from \emph{Planck}, future CMB probes may falsify the currently best inflationary scenarios. We shall conclude in Section~\ref{sec:conclusions}.

\section{Basic definitions} 
\label{sec:basic}
The power spectrum of the primordial curvature perturbation, $\zeta$, seeding structure formation in the universe is defined as 
\be 
\vev{\zeta_{\vec k} \zeta_{\vec{p}}}=(2\pi)^3 \delta^{(3)}(\vec{p}+\vec{k}) P_\zeta (k)~,
\ee
where the dimensionless amplitude of primordial perturbations $\Delta_\zeta(k)$ is defined through 
\be 
P_\zeta(k) = {2\pi^2\over k^3} \Delta^2_\zeta (k).
\ee
The scale dependence of $\Delta^2_\zeta(k)$ is parametrized by the spectral index:
\be 
n_s=1+{\d \ln \Delta^2_\zeta \over \d \ln k}~.
\ee
Likewise, one can also define the scale dependence of the spectral index, which is called the running, as  
\be
\alpha_s\equiv \frac{\d n_s}{\d\ln k}\,, 
\ee
as well as the running of the running, defined as 
\be
\beta_s\equiv{\d \alpha_s\over \d \ln k}\,.
\ee
In all these definitions, it is understood that quantities are evaluated at horizon exit $k_*=aH$ ($k_* = 0.05 \, \rm{Mpc}^{-1}$ throughout this study). In terms of the above parameters, the primordial power spectrum reads
\be
\Delta_\zeta^2(k)= \Delta_\zeta^2(k_\ast) \( k \over k_\ast\)^{n_s -1 +\frac12 \alpha_s \ln(k/k_\ast)+\frac1{3!}\beta_s\ln^2(k/k_\ast)}~.
\ee
In the context of slow-roll, one can have a general idea about the magnitude of the above inflationary parameters in terms of the number of e-folds $N$. If we consider the empirical relation \cite{Barranco:2014ira, Boubekeur:2014xva, Garcia-Bellido:2014gna} $n_s-1\propto 1/N$, one expects that  
\be
\alpha_s\sim {1\over N^2}\lesssim 10^{-4}~~ \textrm{and~~} \beta_s\sim {1\over N^3}\lesssim 10^{-5}\,, 
\ee
for typical choices of the number of e-foldings $N=50-60$. 
The latest {\sl Planck} 2015 temperature and polarization TT,TE,EE+lowP~\cite{planck} data analyses with $r=0$ provide the following constraints:
\begin{align*}
\begin{tabular}{lcl}
$n_s = 0.9586\pm 0.0056$~,\\ \\
$\alpha_s = 0.009 \pm 0.010$~,\\ \\
$\beta_s = 0.025\pm 0.013$~.\\ \\
\end{tabular}
\end{align*}
What is interesting to notice in these constraints, is a slight preference for a positive $\beta_s\sim 10^{-2}$, while as we will explain shortly, slow-roll inflation predicts typically a smaller and {\it negative} $\beta_s$.

The tensor contribution to the primordial power spectrum is parametrized by the tensor-to-scalar ratio $r$
\be
r=P_t(k_\ast)/P_\zeta(k_\ast)~,
\ee
where $P_t(k) \equiv {2\pi^2\over k^3} \Delta^2_t (k)$ is the tensor power spectrum, and it is parametrized at first order as
\be
 \Delta^2_t (k)= \Delta^2_t (k_*)\( k \over k_\ast\)^{n_t}~,
\ee
in which $n_t$ is the spectral index of tensor modes. In the slow-roll regime, the magnitude of $r$ can vary within a large range, and this is the main difficulty in testing inflation through the  detection of $B$-modes. This can be understood in the context of phenomenological parametrizations of inflation \cite{Barranco:2014ira, Boubekeur:2014xva, Garcia-Bellido:2014gna}. In such approaches, the $(n_s, r)$ plane appears to be unevenly filled, and one can even argue on the existence of a ``forbidden zone"~\footnote{This observation has been made previously in different contexts in \cite{Efstathiou:2006ak, Bird:2008cp, Alabidi:2006fu}.}, in the $r$-direction, depending on the precise value of $n_s$, see Figure~\ref{fig:rns}. Future CMB missions aim to reach the important theoretical milestone of $r=2\times 10^{-3}\cdot (60/N)^2$\cite{Bouchet:2011ck}, which would signal super-Planckian inflaton excursions \cite{Lyth:1996im, Boubekeur:2005zm, Boubekeur:2012xn}. 

\section{Most favoured inflationary scenarios}
\label{sec:models}
In the following, we shall review the most favoured models (including their predictions for the different inflationary observables: $r$, $n_t$, $n_s$, $\alpha_s$ and $\beta_s$) after {\sl Planck 2015} data release.

\subsection{Quadratic scenarios}
This class of scenarios represents the simplest theoretical possibility. It includes:\\ \\
{\bf The chaotic scenario}, $V\propto\phi^2$, both with minimal and non-minimal coupling to gravity~\cite{Linde:1983gd, Salopek:1988qh, Futamase:1987ua, Fakir:1990eg, Komatsu:1999mt, Hertzberg:2010dc, Linde:2011nh}. The former is disfavoured with respect to the latter so the non-minimally coupled version is perfectly compatible with current data \cite{Boubekeur:2015xza}. The predictions in the $(n_s, r)$, $(n_s,\alpha_s)$, $(n_s, \beta_s)$ and $(n_t, r)$ planes for these two models ($\phi^2$ and $\xi \mathcal{R}\phi^2$) are depicted in Figures~\ref{fig:rns}, \ref{fig:ans}, \ref{fig:bns} and \ref{fig:nt} for two possible choices for the number of e-folds, $N=50$ and $N=60$~\footnote{The value of $\xi$ ranges from $\xi=0$ to $\xi=0.0065$ in Figures~\ref{fig:rns}, \ref{fig:ans} and \ref{fig:bns}.}.  Notice, from Figure~\ref{fig:rns}, that the trajectories in the $(n_s, r)$ plane for the non-minimally coupled case ($\xi \mathcal{R}\phi^2$) start  always at the point corresponding to the $\phi^2$ model predictions~\footnote{The case of $\xi=0$ is equivalent to the standard inflationary chaotic scenario in which the predictions are $n_s= 1-2/N$ and $r=8/N$, 
 corresponding to $n_s=0.967$ and $r=0.13$, respectively, for $N=60$.}, and then, as the coupling $\xi$ takes positive values, the tensor contribution is reduced, and the scalar spectral index $n_s$ is pushed below scale invariance, see Ref.~\cite{Boubekeur:2015xza}. Negative values of the coupling $\xi$ (not illustrated here) are highly disfavoured by current CMB observations, since they will lead to large values of the tensor-to-scalar ratio $r$. Concerning the running of the scalar spectral index $\alpha_s$, the trajectories for the two quadratic scenarios considered here are depicted in Figure~\ref{fig:ans}. Notice that positive values of the coupling $\xi$ will change the predicted value of $\alpha_s$ in the $\phi^2$ scenario ($\alpha_s=-2/N^2$, corresponding to $\alpha_s=-0.00056$ for $N=60$) to slightly larger values, albeit the trajectories always stay in the $\alpha_s<0$ sub-plane. The running of the running parameter, $\beta_{s}$, barely changes with respect to its predicted value in the non-minimally coupled case (i.e. $\xi=0$, for which $\beta_s=-4/N^3$, giving $\beta_s\simeq -1.8\times 10^{-5}$ for $N=60$) as the coupling $\xi$ gets positive values, see Figure~\ref{fig:bns}. Finally, in Figure~\ref{fig:nt} we see that all models follow the theoretical curve $n_t=-r/8$. In particular, the chaotic $\phi^2$ model predicts a tensor spectral index of $n_t\simeq -0.019$ ($n_t\simeq -0.016$) for $N=50$ ($N=60$); an increasing positive value of $\xi$, within the non-minimally coupled model, diminishes the predicted value to $n_t\simeq-0.018$ ($n_t\simeq-0.010$) for $\xi\simeq 0.0059$ and $N=50$ ($N=60$).

{\bf The Natural inflation scenario} (minimally coupled to gravity), where the inflaton is a Pseudo-Nambu-Goldstone-Boson (PNGB), which potential is invariant under the shift $\phi\to \phi+2 \pi f$, and it is given by 
\be
V(\phi)=V_0\[1 -\cos(\phi/f)\]\,,
\ee
with $f$ the PNGB decay constant~\cite{Freese:1990rb, Adams:1992bn, Kim:2004rp}. It is straightforward to perform the slow-roll analysis and obtain the analytical expressions of the spectral index and the tensor-to-scalar ratio:
\be
\begin{aligned}
n_s & =1-\alpha\[\frac{e^{\alpha N}\(1+\alpha/2\)+1}{e^{\alpha N}\(1+\alpha/2\)-1}\]\,,\\
r & =\frac{8\alpha}{e^{\alpha N}\(1+\alpha/2\)-1}\,,
\end{aligned}
\ee
where the parameter $\alpha$ is defined as~\footnote{$M_{\rm pl}=1/\sqrt{8\pi G_N}\simeq 2.43\times 10^{18}$ GeV is the reduced Planck mass.} $\alpha\equiv M^2_{\rm pl} /f^2$. Notice that for small $\alpha$ (i.e. very large values of $f$) the predictions of the natural inflation scenario coincide with those of the minimally coupled chaotic inflation model $V\propto \phi^2$.  Even if the flatness of the PNGB potential is protected by the shift symmetry, it is not clear whether this structure can be UV completed. For a recent discussion on the issue and some solutions see \eg \cite{Conlon:2012tz, Boubekeur:2013kga, delaFuente:2014aca, Heidenreich:2015wga, Brown:2015lia}.

Figures~\ref{fig:rns}, \ref{fig:ans}, \ref{fig:bns} and \ref{fig:nt} show the predicted trajectories in the $(n_s, r)$, $(n_s,\alpha_s)$, $(n_s, \beta_s)$ and $(n_t, r)$ planes for $N=50$ and $N=60$, and $f$ varying from $3.45 M_{\rm pl}$ to $100 M_{\rm pl}$. For the smallest value of $f$ considered here, $f=3.45 M_{\rm pl}$, a very small value of $n_s\simeq 0.9152$ is found. Agreement with \emph{Planck} data implies that the decay constant satisfies $f > 5.3M_{\rm pl}$, for $N=50-60$. Larger values of $f$ increase the value of the tensor-to-scalar ratio, until the prediction reaches the one of minimal chaotic inflation, as shown in Figure~\ref{fig:rns}. In Figure~\ref{fig:ans}, we illustrate that large values of $f$ lead to small values for the running of the spectral index, which eventually will reach the predictions for the minimal chaotic scenario. In contrast, the value of $\beta_s$, barely changes when $f$ varies, remaining around in $\beta_s\simeq-3\times10^{-5}$ and $\beta_s\simeq-1.7\times10^{-5}$ for $N=50$ and $N=60$, respectively, see Figure~\ref{fig:bns}. Concerning the tensor spectral index, for a value of $f=100M_{\text{pl}}$ the predictions coincide with those of the $\phi^2$ model. Whereas lower value of $f$, corresponds to smaller values $n_t$. For instance,  $n_t\simeq-0.0006$ ($n_t\simeq-0.0002$) for $f=3.45M_{\text{pl}}$ and $N=50$ ($N=60$), following the consistency relation $n_t=-r/8$, as expected (see Figure~\ref{fig:nt}).

\subsection{Higgs-like scenarios} 

This class of scenarios is described by a symmetry breaking potential, 
\be
V(\phi)=\frac{\lambda\(\phi^2-v^2\)^2}{4\(1+\xi\phi^2/M^2_{\rm pl}\)^2}\,,
\ee
alike to the one of the standard model Higgs particle, but with a non-minimal-coupling to the Ricci scalar, $\xi$, see Refs.~\cite{Bezrukov:2007ep, Barvinsky:2008ia, Bezrukov:2009db}. It also includes, as a limiting case (for $\xi\to\infty$), the $R^2$-gravity Starobinsky scenario~\cite{Starobinsky:1980te}. Notice as well that the limiting case $\xi\to0$ corresponds to the quartic potential scenario, $V\propto\phi^4$. One can find a suitable set of inflaton potentials for different values of the inflaton vacuum expectation value $v$ \cite{Linde:2011nh}. In this work we illustrate the predictions of a Higgs-like scenario for $v=0$ and for different positive values of $\xi$, as well as for $N=50$ and $N=60$ e-folds~\footnote{The coupling $\lambda$ cancels out in the slow-roll calculations.}. Figure~\ref{fig:rns}  clearly shows that the limiting case $\xi\to0$, corresponding to the quartic potential $\phi^4$, is not in good agreement with \emph{Planck} data, as its predictions for the inflationary parameters ($n_s\simeq0.941$, $r\simeq0.31$ and $n_s\simeq0.951$, $r\simeq0.26$ for $N=50$ and $N=60$, respectively) are highly disfavoured. When the non-minimal coupling to gravity, $\xi$, is increased, the tensor contribution is reduced, while the predictions reach those corresponding to the Starobinsky scenario, as long as  $\xi>10^2$. In this limit, $n_s\simeq0.961$, $r\simeq0.0041$ ($n_s\simeq0.968$, $r\simeq0.0023$) for $N=50$ ($N=60$), values which are in excellent agreement with current CMB data.

Concerning the running of the spectral index, increasing the value of $\xi$ will drive the values of $\alpha_s$ from the one corresponding to the quartic potential to slightly larger ones, corresponding to the Starobinsky scenario, keeping always the trajectory in the $\alpha_s<0$ sub-plane (see Figure~\ref{fig:ans}). The predictions of the running of the spectral index for the quartic (Starobinsky) scenarios are $\alpha_s\simeq-0.0011$ ($\alpha_s\simeq-0.00074$) for $N=50$, and $\alpha_s\simeq-0.0008$ ($\alpha_s\simeq-0.00052$)  for $N=60$. As in the case of the previous models, the running of running of the spectral index, $\beta_s$, remains almost constant as $\xi$ is varied, as shown in Figure~\ref{fig:bns}. In particular, in the Higgs-like scenario, $\beta_s\simeq-3.5\times10^{-5}$ ($\beta_s\simeq-2.5\times10^{-5}$) for $N=50$ ($N=60$). This model allows for a wide range of values for the tensor spectral index, starting from the predictions from the $\phi^4$ model around $n_t\simeq-0.039$ ($n_t\simeq-0.039$). Then, an increasing value of $\xi$ pushes down the predictions for $n_t$ down to very small values around $n_t\simeq-0.0005$ ($n_t\simeq-0.0003$) for $N=50$ ($N=60$), (and thus coinciding with the values predicted from Starobinsky inflation), along the theoretical curve $n_t=-r/8$ depicted in Figure~\ref{fig:nt}.

\subsection{Hilltop scenarios}
For completeness, we should also consider this class of scenarios, described by potentials 
\be 
V(\phi)=V_0\[1-(\phi/\mu)^p\]\,,
\ee 
since its predictions in the $(n_s, r)$ plane lie very close to the ones associated to the models discussed before~\cite{Boubekeur:2005zm}. Within these scenarios, we can distinguish two sub-cases: 
\begin{enumerate}
\item $p=2$, corresponding to the quadratic hilltop scenario, where inflation takes place close to a local maximum; $V'(\phi)=0$ and $V''(\phi)<0$. 
\item $p>2$,  corresponding to a generalization of the simplest quadratic case, where here inflation happens close to a local maximum where additionally, higher derivatives of the potential vanish, \ie $V'(\phi)=V''(\phi)=V'(\phi)=\cdots =V^{(p-1)}(\phi)=0$ and, again,  $V^{(p)}(\phi)<0$. 
\end{enumerate}

We restrict our analysis to the first case, $p=2$,  in which the spectral index and the tensor-to-scalar ratio read as
\be
\begin{aligned}
n_s & =1-4|\eta_0|\\
r & =2\(1-n_s\)^2e^{N\(n_s-1\)}|\eta_0|^{-1}\,,
\end{aligned}
\ee
with $|\eta_0|=\mu^{-2}M^2_{\rm pl}$. In Figure~\ref{fig:rns} we depict the predictions for this model in the plane $\(n_s,r\)$. The parameter $\eta_0$ varies from $\eta_0=10^{-4}$ to $\eta_0\simeq2\times10^{-2}$, pushing $n_s$ to smaller values as $\eta$ decreases. With $\eta_0\simeq8\times10^{-3}$ we obtain a tensor-to-scalar ratio of $r=0.0375$ for the case N=60, and $r\simeq0.0516$ for $N=50$, both corresponding to a spectral index $n_s\simeq0.968$. Notice from Figures~\ref{fig:ans} and \ref{fig:bns} that for the same value of $\eta_0\simeq8\times10^{-3}$ we obtain a running of the spectral index of $\alpha_s\simeq-0.00107$ and a running of the running $\beta_s\simeq-0.000065$  ($\alpha_s\simeq-0.00073$ and $\beta_s\simeq-0.0000386$) for $N=50$ ($N=60$). In Figure~\ref{fig:nt} we observe that this scenario predicts almost negligible values of the tensor spectral index for the range of values of $\eta_0$ commented above. The predictions reach the smallest values of $n_t$ found in this work: $n_t\simeq-0.0003$ ($n_t\simeq-0.00006$) for $N=50$ ($N=60$) and $\eta_0\simeq2\times10^{-2}$.

\section{Current constraints}

\label{sec:current}

\subsection{Cosmological data and methodology}
We consider the new data on CMB temperature and polarization measured by the \emph{Planck} satellite~\cite{Ade:2015xua,Adam:2015rua,Aghanim:2015wva}. We use the \emph{Planck} TT temperature-only likelihood (hereafter \textit{Planck} TT) and the \emph{Planck}  TT,TE, and EE power spectra data (hereafter \textit{Planck} TTTEEE) up to a maximum multipole number of $\ell_{\mathrm{max}} = 2500$ combined with the \emph{Planck} low-$\ell$ multipole likelihood that extends from $\ell=2$ to $\ell=29$ (denoted as lowP). We use the Boltzmann code CAMB~\cite{Lewis:1999bs} and generate MCMC chains using the publicly available package cosmomc~\cite{Lewis:2002ah}. We consider a $\Lambda$CDM extended model, described by the following set of parameters: 
\begin{equation}\label{parameterPPS}
\{\omega_b,\omega_c, \Theta_s, \tau,\ln{(10^{10} A_{s})},n_s,r, \alpha_s, \beta_s\}~.
\end{equation}
In Table~\ref{tab:priors}, the uniform priors considered on the different cosmological parameters are specified. We do not consider the spectral index for tensor perturbations $n_t$ as an additional parameter in our MCMC analyses, since, as recently shown in \cite{Cabass:2015jwe}, the current and future error bars on this parameter are considerably larger than the predictions of the different theoretical scenarios explored here. Therefore, the tensor spectral index is fixed in what follows to the slow-roll consistency relation value, $n_t = -r/8$.

\begin{table}
\begin{center}
\begin{tabular}{ccc}
\hline\hline
 Parameter & Physical Meaning  & Prior\\
\hline
$\omega_b\equiv\Omega_{b}h^2$ & Baryon density &$0.005 \to 0.1$\\
$\omega_c\equiv\Omega_{c}h^2$ &Cold dark matter density &$0.01 \to 0.99$\\
$\Theta_s$ & Angular scale of recombination & $0.5 \to 10$\\ 
$\tau$ &Reionization optical depth &$0.01 \to 0.8$\\
$\ln{(10^{10} A_{s})}$& Primordial scalar amplitude& $2.7 \to 4$\\
$n_{s}$ &Scalar spectral index  &$0.9 \to 1.1$\\
$\alpha_s$   &Running of $n_s$ &    $-0.04 \to 0.06$\\
$\beta_s$    &Running of $\alpha_s$ &    $-0.04 \to 0.08$\\
$r$               &Tensor-to-scalar ratio&     $0 \to 2$\\
\hline\hline
\end{tabular}
\caption{Uniform priors for the cosmological parameters considered in the present analysis.}
\label{tab:priors}
\end{center}
\end{table}

\begin{table*}
\begin{center}
\begin{tabular}{ccc}
\hline \hline
\rule[-2mm]{0mm}{6mm}

Parameter & \textit{Planck} TT+lowP & \textit{Planck} TT,TE,EE+lowP\\
\hline
\hline
$r$ ($95\%$ CL)  &$<0.27$&$<0.23$\\ 

$n_s$  &$0.959 \pm 0.008 $&$0.9591 \pm 0.0056$\\ 

$\alpha_s$  &$0.0081 \pm 0.014 $&$0.0077 \pm 0.011$\\ 

$\beta_s$  &$0.034 \pm 0.016 $&$0.0313 \pm 0.014 $\\ 
\hline
\hline
\end{tabular}
\caption{$95\%$~CL constraints on the tensor-to-scalar-ratio $r$ and mean values (together with their $68\%$~CL errors) of $n_s$, $\alpha_s$ and $\beta_s$ obtained with the two possible data combinations considered in this study.}
\label{tab:outbounds}
\end{center}
\end{table*}

\subsection{Results}
While the latest \textit{Planck} data provide evidence against some of the models explored here~\cite{planck}, these measurements can not single out the responsible mechanism for the inflationary process, nor to falsify this theoretical scenario by themselves. 

This can be noticed from the contours shown in Figures~\ref{fig:rns} \ref{fig:ans} and \ref{fig:bns}, where it is clear that all the models described above have some trajectories in the  $(n_s, r)$, $(n_s,\alpha_s)$ and $(n_s, \beta_s)$ planes which lie within the current $68\%$ and/or $95\%$~CL allowed regions. Figure \ref{fig:rns} depicts the current $68\%$ and $95\%$~CL allowed contours in the $(n_s, r)$ plane from  {\sl Planck} TT plus lowP data, as well as from {\sl Planck} TT plus lowP data plus TTEETE measurements, together with the predictions from Natural, Hilltop, Higgs-like, quartic, chaotic~\footnote{The chaotic model is studied both in its minimally and non-minimally coupled versions.} and Starobinsky inflationary scenarios, for both $N=50$ and $N=60$ e-folds. The addition of EE and TE spectra to {\sl Planck}  TT plus lowP data helps in constraining the scalar spectral index $n_s$, however there is only a mild  improvement in the tensor-to-scalar ratio upper bound. Notice, as previously stated, that the predictions for the inflationary parameters $n_s$ and $r$ from these models are all well within the current $68\%$ and/or $95\%$~CL allowed regions and therefore all of them (except for the case of the $\phi^4$ potential with $N=50$) are still feasible. One could ask if current measurements of other inflationary parameters, as the running of the scalar spectral index $\alpha_s$ and/or its running, $\beta_s$, may help in disentangling among the plethora of models still allowed by current data. Figure~\ref{fig:ans}, illustrates, together with the trajectories in the $(n_s,\alpha_s)$ plane for the models explored here, the $68\%$ and $95\%$~CL allowed regions from {\sl Planck} TT plus lowP data as well as from TTEETE plus lowP measurements. Notice that current bounds on $\alpha_s$ are unable to discard any of the possible inflationary models. Figure~\ref{fig:bns} shows the equivalent but in the  $(n_s,\beta_s)$ plane. Interestingly, {\sl Planck} measurements of $\beta_s$ seem to exclude the value $\beta_s=0$ at the $\sim 2\sigma$ level. The theoretical scenarios illustrated here could be ruled out with a much higher significance if the value of $\beta_s$ preferred by {\sl Planck} 2015 measurements (i.e. $\beta\simeq 0.025$) is confirmed by future CMB data. We shall explore this possibility in the next section. 

Table~\ref{tab:outbounds} shows the $95\%$~CL bounds on the tensor-to-scalar-ratio $r$ as well as the mean values and $68\%$~CL errors of the remaining inflationary parameters $n_s$, $\alpha_s$ and $\beta_s$ obtained with the two possible data combinations considered in this study. Notice that the limits on $r$ are considerably relaxed when adding the running and the running of the running as additional parameters in the analyses. The mean values and the errors on $n_s$ and $\beta_s$ are in very good agreement with those found by the \emph{Planck} collaboration and reported in Ref.~\cite{planck}. 
 
\section{Forecasts}
\label{sec:future}

The aim of this section is to forecast the potential of future CMB satellites in constraining the $\left\{r,n_s,\alpha_s,\beta_s \right\}$ parameter space via the Fisher matrix formalism.

\subsection{CMB Likelihood}
Assuming that the fraction of sky  surveyed $f_{\textrm{sky}}$ is the same for CMB temperature and polarization measurements, the likelihood associated to a single frequency CMB experiment can be written as 
\begin{eqnarray}\label{Eq:Likelihood}
-2 \, \ln \, \mathcal{L}^{\rm{CMB}} &=& \sum_{\ell} (2\ell + 1)f_{sky} \left[ \ln\left( \frac{C_\ell ^{BB}}{\hat{C}_\ell^{BB}} \right) - \frac{\hat{C}_\ell^{BB}}{C_\ell ^{BB}}   \right. \nonumber \\
&+&  \left. \ln\left( \frac{C_\ell ^{TT}C_\ell ^{EE} -(C_\ell ^{TE})^2}{\hat{C}_\ell ^{TT}\hat{C}_\ell ^{EE} -(\hat{C}_\ell ^{TE})^2 } \right) -3  \right.  \\ \nonumber
&+&  \left. \frac{\hat{C}_\ell ^{TT}C_\ell ^{EE}+C_\ell^{TT} \hat{C}_\ell^{EE}  -2 C_\ell ^{TE} \hat{C}_\ell ^{TE}}{C_\ell ^{TT}C_\ell ^{EE} -(C_\ell ^{TE})^2 } \right]~,
\end{eqnarray}
where the $C_\ell^{XY}$ ($\hat{C}_\ell^{XY}$) refer to the theoretical (measured) power spectra for $X, Y = {T, E, B}$. Due to the finite resolution of the spectra, there will be an induced noise in the map that should be added to the $C_\ell$. In addition, following~\cite{Verde:2005ff} we will also include the foreground contribution to the map as a residual noise, and therefore
\begin{eqnarray}\label{Eq:Clcontribution}
C_\ell = C^{th}_\ell + N_\ell+ R^F_\ell~,
\end{eqnarray}
where $C^{th}_\ell$ will be our theoretical power spectra (computed by the Boltzmann solver codes CAMB~\cite{Lewis:1999bs} or CLASS~\cite{Lesgourgues:2011re}), $N_\ell$ is the instrumental noise (which is a function of the frequency channel, see below) and $R^F_\ell$ refers to the residual foreground subtraction (which will also depend on the frequency channel). This latter quantity reads as
\begin{eqnarray}\label{Eq:ForegroundCl}
R^F_\ell (\nu)&=& \sum_i ^{N_{fore}} \left\{\sigma_i(\nu) C^i_{\ell}(\nu) \right.  \\ \nonumber
&+& \left. N_{\ell}(\nu) \frac{4}{N_{chan}(N_{chan}-1)} \frac{C^i_{\ell}(\nu)}{C^i_{\ell}(\nu_F)} \right\}~,
\end{eqnarray}
where the first term corresponds to the uncertainty of a given foreground at a given frequency $\nu$, $C^i_{\ell}(\nu)$ and  $\sigma_i(\nu)$ represent the power spectra and the foreground subtraction level, respectively. The second term in Eq.~(\ref{Eq:ForegroundCl}) takes into account for the instrumental noise of the channel at which the foreground model is constructed, and $\nu_F$ is the frequency at which the foreground is modelled. In the case of a multifrequency experiment, as Planck or COrE, the expression for the likelihood Eq.~\eqref{Eq:Likelihood} still holds. However, in such a scenario, the total noise power that should be added to the $C_\ell$ is written in terms of a weighted combination of the noises from the different channels~\cite{Verde:2005ff}. Therefore, for a multifrequency experiment, Eq.~(\ref{Eq:Clcontribution}) reads as
\begin{eqnarray}\label{Eq:Clcontribution2}
C_\ell = C^{th}_\ell + N_\ell^{\rm{eff}}~,
\end{eqnarray}
where the effective noise term is given by
\begin{eqnarray}\label{Eq:N_eff}
(N_\ell^{\rm{eff}})^{-2} &=& \sum_{i,j \geq  i} ^{N_{chan}} \left[ \left(R^F_\ell(\nu_i) + N_\ell(\nu_i) \right) \right. \times  \\ \nonumber 
&&   \left( R^F_\ell(\nu_j) + N_\ell(\nu_j) \right) \frac{1}{2} \left( 1 + \delta_{ij} \right) ]^{-1}~.
\end{eqnarray}

We focus here on the future satellite experiment COrE~\cite{Bouchet:2011ck}, covering $70\%$ of the sky. In the next sections we will describe the modelling of the experimental resolution and the main foregrounds for this future CMB mission, and therefore in what follows the numbers quoted will always refer to $f_{\textrm{sky}}=0.7$. 

\subsubsection{Instrumental Noise}

The sensitivity of the detectors of a given CMB experiment is finite; thus, a certain noise will be induced in the map due to the deconvolution of a Gaussian beam, which reads as~\cite{Knox:1995dq}
\begin{equation}\label{Eq:StatNoise}
N_\ell^{XY} = \sigma^X \sigma^Y \delta_{XY} \exp \left[ \ell \left( \ell+1 \right) \frac{\theta^2}{8\hspace{0.4ex}{\rm ln}2} \right]~,
\end{equation}
where $\sigma^X$ corresponds to the temperature and polarization sensitivity of the channel, respectively ($X =\{T, P\}$), and $\theta$ is the Full Width at Half Maximum (FWHM) of the beam. We follow here the specifications for the future COrE mission given in Ref.~\cite{Bouchet:2011ck}, see Table~\ref{Table:Experiments} of Appendix B.

\subsubsection{Foregrounds}
Foregrounds, consisting of radio emissions from the galaxy and/or other sources at the same frequency to that of the CMB signature, will clearly be the dominant limiting factors in extracting the cosmological information from the maps. 
In the case of the polarized signal, foregrounds are critical as they are orders of magnitude higher than the primordial signal in some cases. The usual strategy followed to deal with the foregrounds is to exploit their spectral dependence. Several recent works~\cite{Ade:2015fwj,Ade:2015tva,Ade:2015qkp,Adam:2014bub} have shown that an accurate multifrequency approach to correctly handle foregrounds is mandatory. Here we will briefly discuss the physical origin of the main foregrounds relevant for the COrE mission~\footnote{Other two sources of foregrounds are the Anomalous Microwave Emission and the Free-Free emission (see Ref.~\cite{Ade:2014zja} for details related to their parametrized power spectra) not discussed here, as their impact at the frequency range of interest is negligible.} and their up-to-date modelling, as provided by the \emph{Planck} team. \\

\paragraph{Synchrotron emission} $ $\\

Synchrotron emission results from the interaction of high energy electrons with the magnetic fields of the galaxy, and its signature will be present in both temperature and polarization maps. Giving the dependence of the synchrotron optical depth with frequency, the power of synchrotron emission $C_{\ell}^S$ grows with decreasing frequency. It is usually modelled using maps at $408$~MHz~\cite{Haslam:1982zz} and with the WMAP K-band at $23$ GHz~\cite{Bennett:2003ca}. The synchrotron power spectra is well fitted using a simple power law for both $\ell$ and $\nu$. The latest \emph{Planck} model~\cite{Ade:2014zja} is
\begin{eqnarray}\label{Eq:Synchrotron}
C_{\ell}^S= A_S  \left( \frac{\ell}{\ell_S} \right)^{\alpha_{S}}  \left( \frac{\nu}{\nu_S} \right)^{2\beta_{S}}~,
\end{eqnarray}
where the values of the different parameters are shown in Table~\ref{tab:Foregroundstab}. \\

\paragraph{Thermal Dust}  $ $\\

Contrarily to synchrotron emission, the power at which thermal dust radiates grows with frequency. \emph{Planck} has modelled the dust contamination using a Modified Black Body for which $T_D = 19.6 \, \rm{K}$. The intensity~\cite{Ade:2014zja}  and polarization~\cite{Adam:2014bub} spectra can be written as
\begin{eqnarray}\label{Eq:Dust}
C_{\ell}^D= A_D \left( \frac{\ell}{\ell_D} \right)^{\alpha_D} \left( \frac{\nu}{\nu_D} \right)^{2\beta_D-4} \left(\frac{B_\nu(T_D)}{B_{\nu_D}(T_D)}\right)^2
\end{eqnarray}
and 
\begin{eqnarray}\label{Eq:DustPol}
C_{\ell \ p}^D= \frac{A_D^p}{2\pi} \left( \frac{\ell}{\ell_D^p} \right)^{\alpha_D^p+2} \left( \frac{\nu}{\nu_D^p} \right)^{2\beta_D^p-4} \left(\frac{B_\nu(T_D)}{B_{\nu_D^p}(T_D)}\right)^2
\end{eqnarray}
respectively, where $B_\nu(T) = 2 h \nu^3c^{-2}/(e^{\frac{h\nu}{kT}}-1)$. The values of the different parameters are specified in Table~\ref{tab:Foregroundstab}.

\begin{table}[t]
\begin{center}
\begin{tabular}{lccc}
\hline\hline
Foreground & Parameter & \emph{Planck} \\
\hline
& $A_S$ ($ \mu K_{\rm{CMB}}^2$) & $(4.2 \pm 0.4)\times10^9$   \\
 & $\nu_S$ (GHz) & 0.408   \\
Synchrotron & $\ell_S$  & $100$   \\
& $\beta_{Syn}$  & $-3.00 \pm 0.05$  \\
& $\alpha_{Syn}$  & $-2.5  \pm 0.02 $ \\
\hline
 & $A_D$ ($ \mu K_{\rm{CMB}}^2$) & $ 40 \pm 3 $\footnote{From Ref.~\cite{Ade:2014zja}, after applying color corrections and conversion units.}   \\
 & $\nu_D$ (GHz) & 353   \\
Dust & $\ell_D$  & $100$   \\
& $\beta_D$  & $1.51 \pm 0.01 $    \\
& $\alpha_D$  & $-2.4  \pm 0.02 $    \\
\hline
 & $A_{EE}$ ($ \mu K_{\rm{CMB}}^2$) & $ 247 \pm 3  $~\footnote{From Table 1 of Ref.~\cite{Adam:2014bub}, after applying the color corrections.}  \\
  & $A_{BB}/A_{EE}$  & $0.53 \pm 0.01$   \\
 & $\nu_D^p$ (GHz) & 353   \\
Dust Polarization & $\ell_D^p$  & $80$   \\
& $\beta_D^p$  & $1.59 \pm 0.17 $    \\
& $\alpha_{EE}$  & $-0.42  \pm 0.02 $   \\
& $\alpha_{BB}$  & $-0.44  \pm 0.03 $   \\ \hline\hline
\end{tabular}
\caption{Parameters for the different foregrounds considered in this study, for both the intensity and polarized emissions. For the intensity signal the models are fitted for $\ell < 100$ and, for polarization, for $60 < \ell < 500$. As commonly carried out in the literature, we will extrapolate the models to higher and lower multipoles for both the intensity and polarization spectra.} 
\label{tab:Foregroundstab}
\end{center}
\end{table}

\subsubsection{Statistical Method}
In order to forecast the errors of the different parameters we follow the widely used Fisher matrix formalism~\cite{Fisher:1935}. The Fisher matrix is defined as the expectation value of the second derivative around the maximum of the likelihood
\begin{equation}
F_{ij} = -\langle  \frac{\partial^2 \mathcal{L}}{\partial \theta_i \partial \theta_j} \rangle |_{ \vec{\theta} = \vec{\theta}_{fid}},
\end{equation}
where $\theta_i$ represent a cosmological parameter, and $\theta_{i,fid}$ represents the fiducial value for the parameter. The Cram\'er-Rao bound ensures that for unbiased estimators the best achievable 1$\sigma$ error for a given parameter marginalized over the other parameters is 
\begin{equation}
\sigma_{\theta_i} =\sqrt{F^{-1} _{ii}}~,
\end{equation}
with $F^{-1}$ the inverse of the Fisher matrix. 

\subsubsection{Foreground removal} 
As argued in the previous section, the main limitation for future CMB observations is the foreground contamination. Among the two polarized foregrounds specified above, the most dangerous one when measuring the tensor-to-scalar ratio $r$ is the galactic dust component, as, in general, it gives the largest contribution at the \emph{Planck} and COrE frequencies. 

The issue of foreground removal is a delicate one. Many techniques like template cleaning, bayesian estimation, internal linear combination or independent component analysis are used for this purpose (see~\cite{Dunkley:2008am} for a summary). For example, in Ref.~\cite{Betoule:2009pq}, a study forecasting errors on $r$ is performed, without any assumption of the properties of the foregrounds. In Refs.~\cite{Errard:2015cxa,Remazeilles:2015hpa} the errors on the different cosmological parameters are obtained after marginalising over the foregrounds following some simple models for their spectra. Here, following the approach of~\cite{Verde:2005ff}, we will assume a simple model for the foregrounds (see  Eqs.~(\ref{Eq:Synchrotron}), (\ref{Eq:Dust}) and (\ref{Eq:DustPol})). We shall also assume in the following, for simplicity, that the foregrounds will be subtracted by a constant amount. Given that the Planck mission has achieved a less than 10\% foreground removal in power, for the COrE mission, due to the high number and the high sensitivity of channels devoted to the study of the dust, one should expect that power could be removed at the 1\% level, which is equivalent to set in Eq.~(\ref{Eq:ForegroundCl}) $\sigma_F(\nu) = 0.01$.

\begin{table*} [t!]
\begin{tabular*}{\textwidth}{{@{\extracolsep{\fill} }cccccc}} 
\toprule
Parameter & Fiducial & \emph{Planck} 2015~\footnote{Using TTTEEE + lowP \emph{Planck} 2015 data.} & \emph{Planck} (Fisher forecast)~\footnote{Using the \emph{Planck} foreground specifications and the $100$, $143$ and $217$ GHz channels.} & COrE $\sigma_F = 0.01$\footnote{Using $\sigma_F = 0.01$ and the 105, 135 165 and 195 GHz channels.}   & COrE $\sigma_F = 0$~\footnote{Using only resolution noise and the 105, 135 165 and 195 GHz channels.} \\
\hline 
$\Omega_{\rm b} h^2$	& 0.02223 	&    0.00017	&    0.00013        	& 0.000065	 &     0.000052   \\ 
$\Omega_{\rm DM} h^2$  & 0.1202		 &   0.0015    	&   0.0012   		& 0.00076 	  &   0.00036  	 \\ 
$h$				        & 0.6762		&    0.0069  	&    0.0054  		& 0.0031 	   	&   0.0014  	 \\ 
$\tau_{\rm reio } $ 		& 0.079		&    0.019	        &   0.018   	  	& 0.0084 	      	&   0.0024  	\\ 
$\ln (10^{10} A_s)$   		& 3.117		 &   0.037 		&   0.036   	 	& 0.015	      	&   0.0044   	\\ 
$n_s$ 				& 0.9591		 &  0.0056 	&  0.0053 	      	  	 & 0.0034         &   0.0023   	\\ 
$\alpha_s $			& 0.0077 		&  0.011 		&  0.0077  		 & 0.0040  	 &   0.0036  \\  
$\beta_s$			 	& 0.0313		 &  0.014 		&  0.019   			&  0.0088 		&  0.0065  \\ 
$r$				 	& 0			&  $< 0.23$ 	&  $< 0.38$  	  	&  $ < 0.016 $ &  $<  0.0001$ \\ 
\hline \hline 
\end{tabular*}
\caption{Results for the Fisher Matrix Analysis, 68\% CL for all parameters and for $r$ at 95\% CL. The fiducial has been assumed to be the same for each run as given in the second column. We have used $\ell^{TT}_{\rm{max}} = 2000$ and $\ell^{BB}_{\rm{max}} = 500$.} 
\label{Table:ForecastComparison}
\end{table*} 

\subsection{Results}

\subsubsection{Future satellite CMB missions}
In the following, we shall apply the Fisher matrix method to the future CMB mission COrE (see Appendix A for a consistency check of our method), although similar results could be obtained for other future CMB satellite experiment. We shall use the $105$, $135$, $165$ and $195$ GHz channels for all the runs, see Table~\ref{Table:Experiments} in the Appendix B~\footnote{For recent CMB forecasts see Refs.~\cite{Creminelli:2015oda,Huang:2015gca}, where however the parameters $\alpha_s$ and $\beta_s$ where not considered.}. 

We perform two different analyses. The first one considers no foreground contamination. The second one relies on a 1\% foreground subtraction in power ($\sigma_F = 0.01$). We assume no \emph{delensing} on the the B-mode signal. The results are shown in Table~\ref{Table:ForecastComparison}. Comparison between the fifth and sixth columns confirm, numerically, the very-well known fact that foregrounds will be the major limitation for future CMB missions when extracting the tensor-to-scalar-ratio $r$. 

\begin{table}
\begin{center}
\begin{tabular}{cccccc}
\toprule 
$r$ 		&	$\sigma_r$	   & $\sigma_{n_s}$ 	&   $\sigma_{\alpha_s}$ 		&   $\sigma_{\beta_s}$	  	\\ \hline
0.1		& 0.0096 			 &  0.0034 			&  0.0040  			 &   0.0086   				\\  
0.01		& 0.0089		        	&  0.0034 				&  0.0040 	        			 &   0.0085 				\\ 
\hline \hline 
\end{tabular}
\caption{Results from the Fisher matrix analysis. The fiducial model corresponds to the \emph{Planck} 2015 best-fit values ($n_s = 0.9591$, $\alpha_s = 0.0077$,  $\beta_s = 0.0313$) and the forecasted errors are obtained assuming a foreground removal $\sigma_F = 0.01$ for the future COrE mission.} 
\label{Table:Uncertainties1}
\end{center}
\end{table} 

\begin{table}
\begin{center}
\begin{tabular}{cccccc}
\toprule 
$r$ 		&	$\sigma_r$	   & $\sigma_{n_s}$ 	&   $\sigma_{\alpha_s}$ 		&   $\sigma_{\beta_s}$	  	\\ \hline
0.1		& 0.0096		 	   &  0.0040 		&  0.0042 	        				 &   0.011 					\\ 
0.01		& 0.0088		 	   &  0.0041 		&  0.0043  				&   0.012  				\\  
\hline \hline 
\end{tabular}
\caption{As Table~\ref{Table:Uncertainties1} but for a fiducial model based on the values predicted by the theoretical models explored here ($n_s = 0.9591$, $\alpha_s = -0.0005$,  $\beta_s = 0$).} 
\label{Table:Uncertainties2}
\end{center}
\end{table}

\subsubsection{Future constraints on inflationary parameters}
\begin{figure*}
\includegraphics[width=0.9\textwidth]{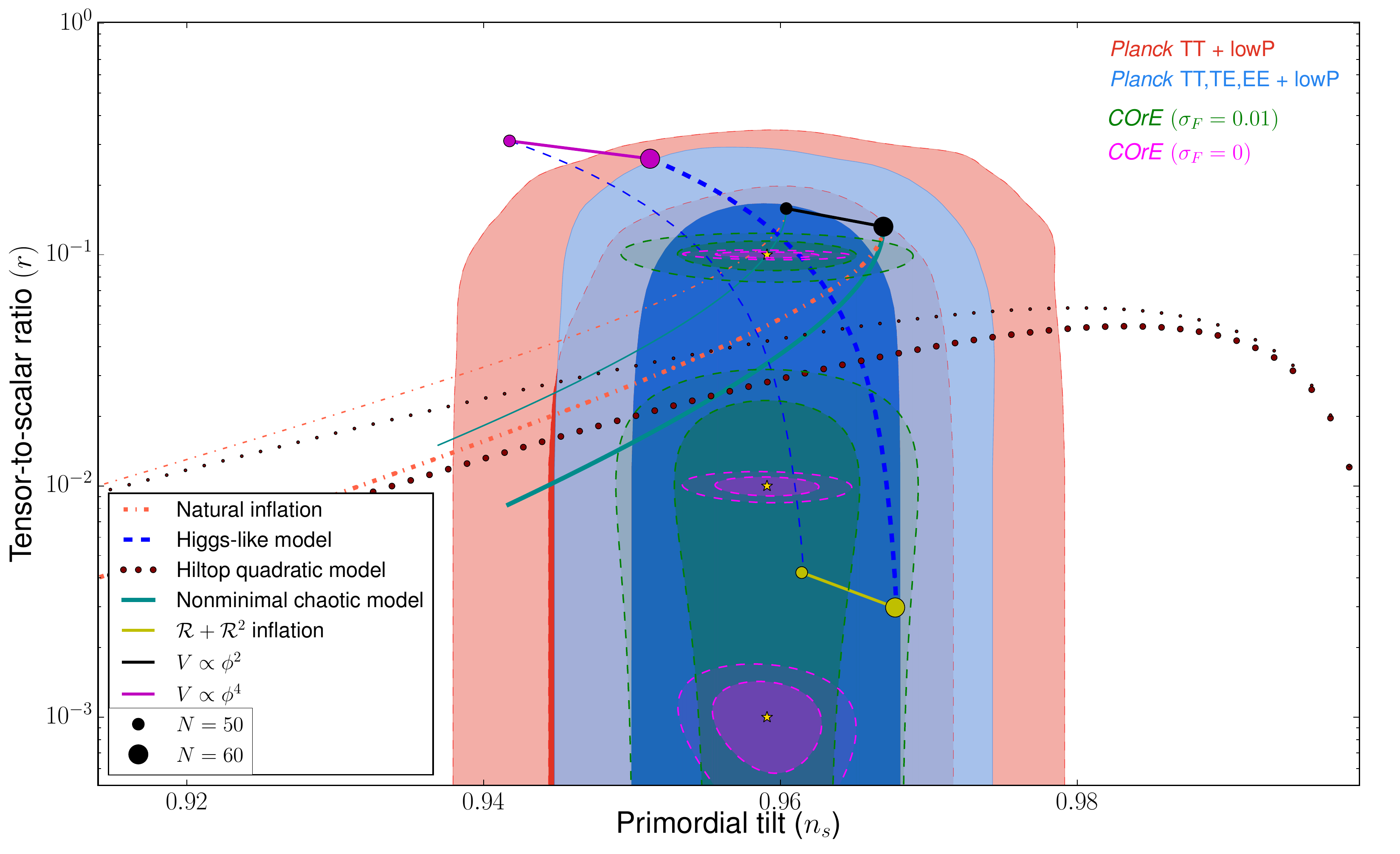} 
\caption{$68\%$ and $95\%$ CL allowed contours from \textit{Planck} TT and lowP and from \textit{Planck} TTTEEE plus lowP in the $(n_s, r)$ plane. We show as well the forecasted $68\%$ and $95\%$ CL contours from the CMB mission COrE considering a 1\% foreground removal in power ($\sigma_F = 0.01$) and perfect foreground subtraction ($\sigma_F = 0$). Notice that if the residual foreground emission is only removed to 1\% level, the future COrE mission may no disentangle among the different models in the $(n_s, r)$ plane. However, the level at which the foregrounds could be removed may be lower than 1\%, but this will only be known once future measurements are performed.}
\label{fig:rns}
\end{figure*}

\begin{figure*}
\includegraphics[width=0.9\textwidth]{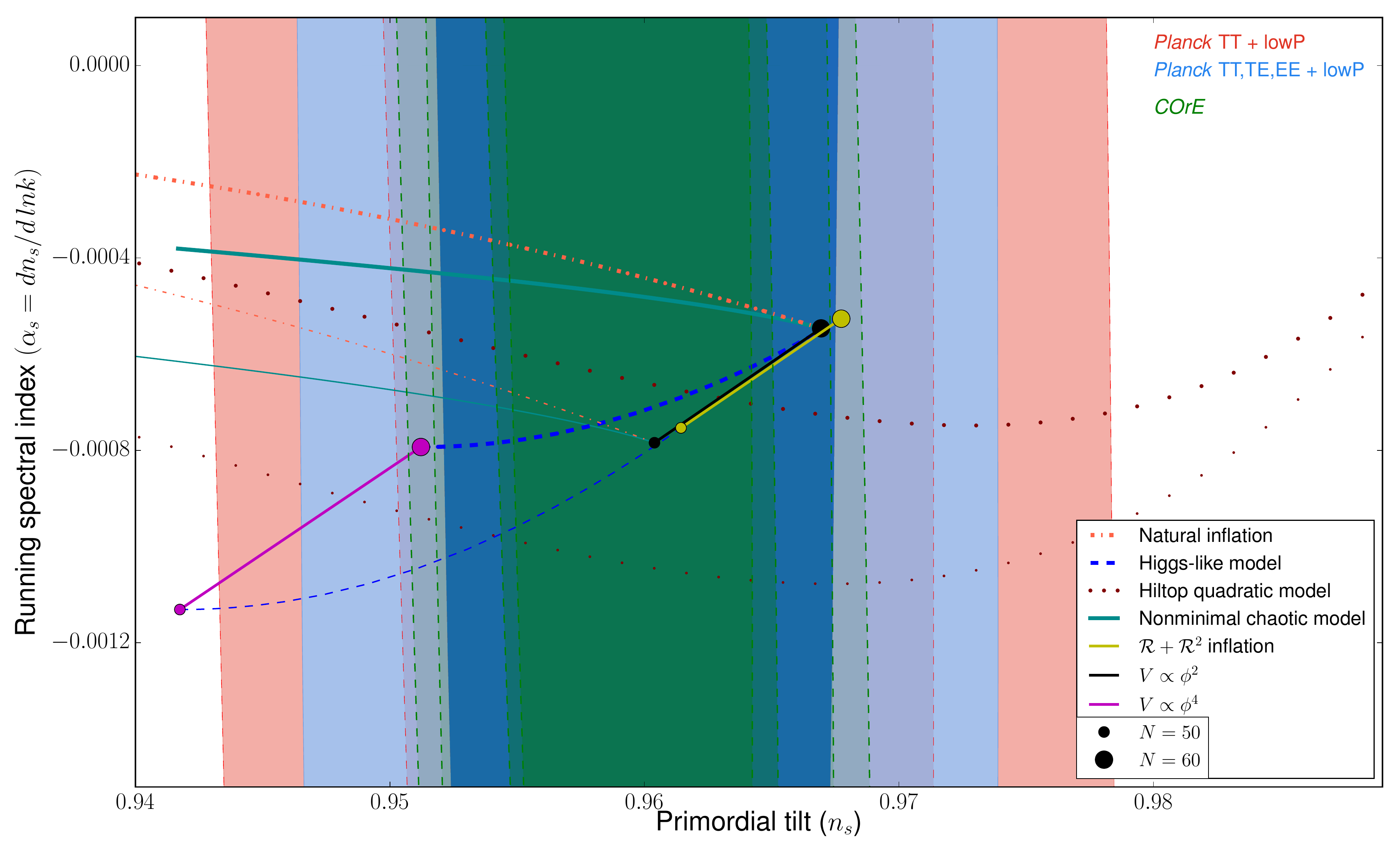} 
\caption{As Figure~\ref{fig:rns} but in the $(n_s, \alpha_s)$ plane, assuming a 1\% foreground subtraction in power.}
\label{fig:ans}
\end{figure*}

\begin{figure*}
\includegraphics[width=0.9\textwidth]{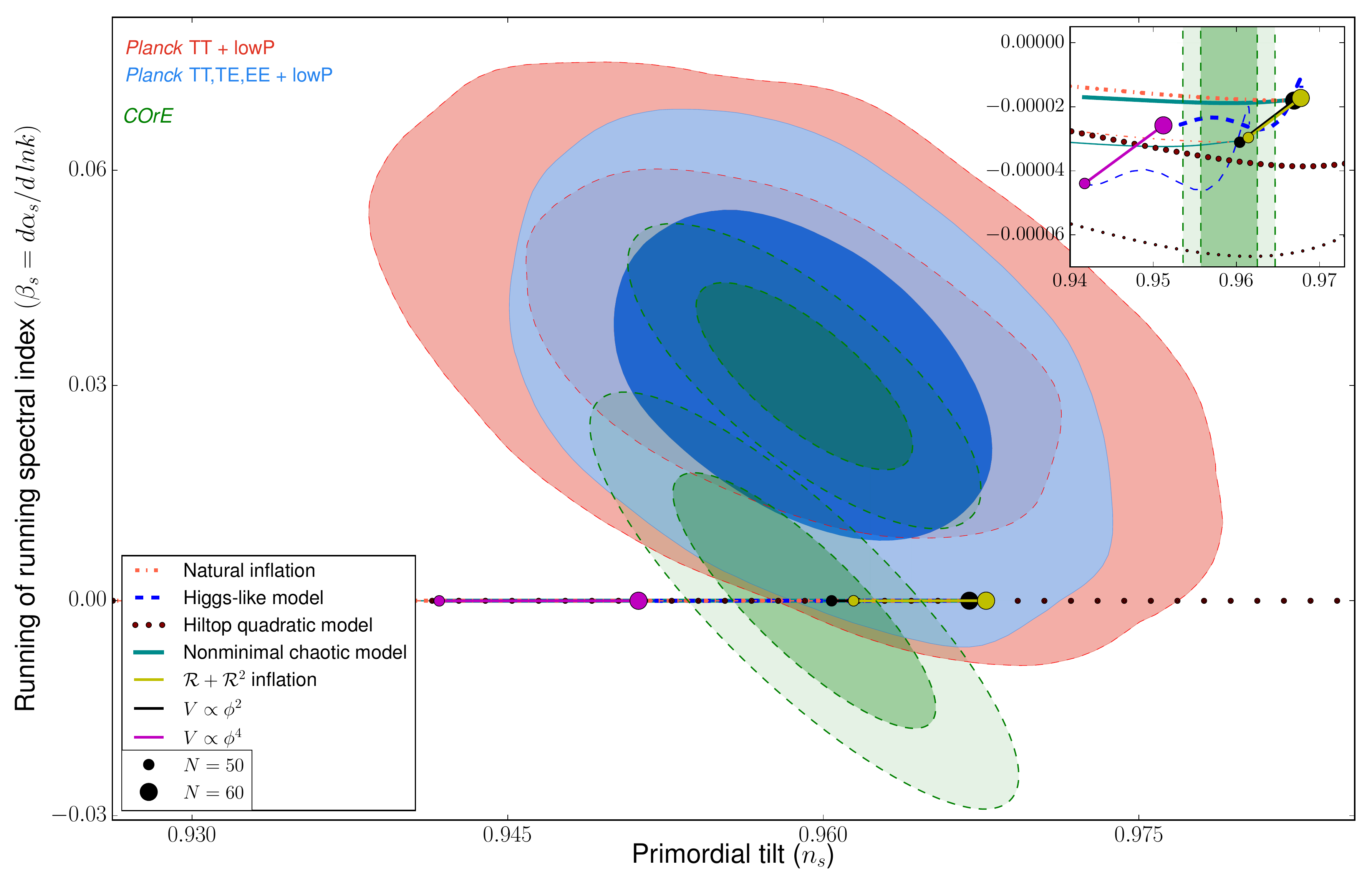} 
\caption{As Figure~\ref{fig:rns} but in the $(n_s, \beta_s)$ plane, assuming a 1\% foreground subtraction in power. If nature has chosen a value of $\beta$ close to the current  best-fit value from \textit{Planck}, then the inflationary models considered here could be excluded by at a high confidence level by future CMB observations.}
\label{fig:bns}
\end{figure*}

Figures~\ref{fig:rns},~\ref{fig:ans} and \ref{fig:bns} show, together with the theoretical predictions and the current constraints from \emph{Planck} measurements, the results of our COrE forecasts for two possible values of the tensor-to-scalar ratio ($r=0.1$ and $0.01$) and two possible fiducial models. The values of the inflationary parameters for the first fiducial model are $n_s = 0.9591$, $\alpha_s = 0.0077$ and $\beta_s = 0.0313$, which correspond to the best-fit to Planck data. For the second fiducial model,  which aims to lie within the region covered by the theoretical models explored here, the values are $n_s = 0.9591$, $\alpha_s = -0.0005$ and $\beta_s = 0$. Tables \ref{Table:Uncertainties1} and \ref{Table:Uncertainties2} show the $1\sigma$ errors on the inflationary parameters for these two fiducial models. Notice that the uncertainties on the $n_s$, $\alpha_s$, and $\beta_s$ barely depend on the fiducial value of $r$, as the tensor-to-scalar ratio is not strongly degenerate with these parameters. The quantity $\sigma_r$ is the expected error from the COrE experiment. Notice that the error is always larger than the COrE sensitivity limit, $\sim 8\times 10^{-3}$ (higher than the target of Ref.~\cite{Creminelli:2015oda}, which was the theoretical milestone of $2 \times 10^{-3}$).  However, the parameter space, the treatment of the foreground removal and the \emph{delensing} assumptions of future CMB missions for the present study and those of Ref.~\cite{Creminelli:2015oda} are different.

\begin{figure*}
\includegraphics[width=0.9\textwidth]{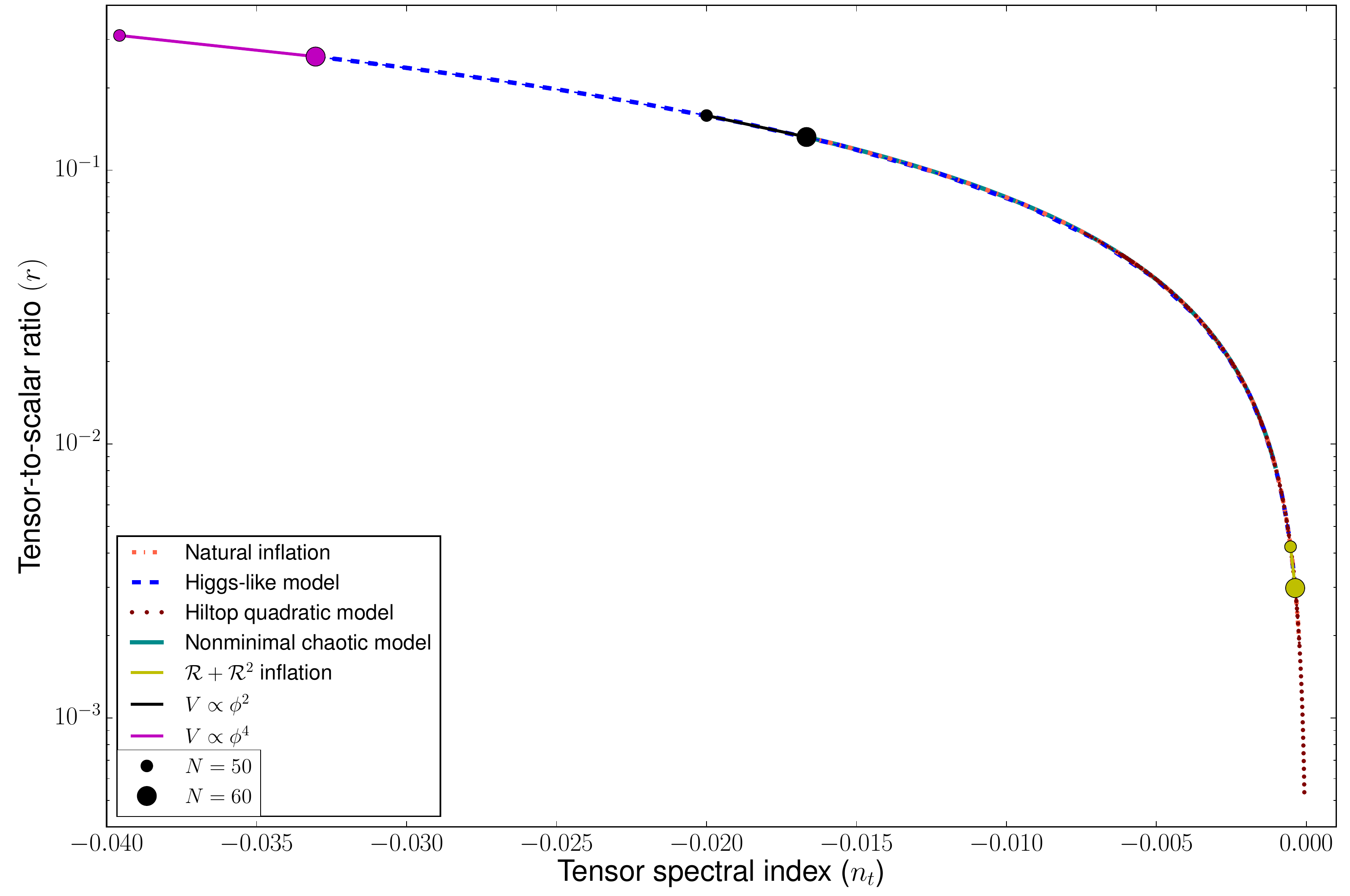} 
\caption{Theoretical predictions in the $(n_t, r)$ plane for the most favoured inflationary models studied here. We do not illustrate the current nor the forecasted contours, as the size of the region depicted here lies well within their $1\sigma$ range, showing that $n_t$ can not help much in distinguishing among these theoretical possibilities (see the recent \cite{Cabass:2015jwe}).}
\label{fig:nt}
\end{figure*}

From the forecasted errors in Tables \ref{Table:Uncertainties1} and \ref{Table:Uncertainties2} and Figure~\ref{fig:rns}  we notice that the measurement of $r$ will not be enough to discriminate between the models on the $(r, n_s)$ plane. As Figures~\ref{fig:ans} and \ref{fig:bns} show, the forecasted errors on the additional parameters considered here, $\alpha_s$ and $\beta_s$, are wider by an order of magnitude or more than the region of values for which the most favoured inflationary models explored in Section \ref{sec:models} spread. Thus, there is no hope in disentangling between the different models using these parameters when the data points to their nominal values of $\alpha_s \approx -0.0005$ and $\beta_s \approx -10^{-5}$. However, notice from Figure~\ref{fig:bns}, that if the best-fit value of the $\beta_s$ parameter arising from future CMB data agrees with its current best-fit from \emph{Planck} measurements, then, this parameter  could allow to exclude the inflationary models explored here at a high confidence level (with the precise significance level depending on the particular model under consideration). Finally (see also Ref.~\cite{Cabass:2015jwe}), the error bars on $n_t$ expected from the CorE mission will be $\sim 3-4$ times larger than the spread of the slow-roll predicted  values (as shown in Figure~\ref{fig:nt}) and therefore this parameter does not help in disentangling among the possible theoretical schemes.
 
\section{Conclusions}
\label{sec:conclusions}

The recent 2015 \emph{Planck} measurements still allow many of the possible theoretical scenarios (as  quadratic-like, Higgs-like and Hilltop models) as the underlying inflationary mechanism. A firm confirmation of the inflationary paradigm would require a detection of the primordial gravitational wave signal. However, in order to single out a theoretical model, the usual two slow-roll parameters, that is, the scalar spectral index $n_s$ and the tensor-to-scalar ratio $r$, may not be sufficient. The reason is due to the fact that the $(n_s, r)$ plane appears to be unevenly filled, with a potentially \emph{forbidden zone} and other highly populated regions in which mostly all the theoretical predictions lie. In this regard, we have explored the discriminating power of two other observables, the running $\alpha_s$ and the running thereof $\beta_s$. Our analyses of \emph{Planck} temperature and polarization data show that the current errors on the former two quantities are large, and therefore they do not help in discarding some of the possibilities, even if the present mean value of $\beta_s$ lies $2 \sigma$ above its predictions in the most favoured inflationary models explored here. However, future CMB measurements, such as the COrE mission, have the potential to rule-out some theoretical possibilities at a much higher significance, provided the best-fit values for these additional parameters do not change significantly from their current estimates. Our forecasts (which rely on both a simple model for foregrounds and assume an \emph{ad-hoc} $1\%$ foreground removal) show that COrE may help enormously in unraveling the inflationary mechanism via its measurement of $\beta_s$, especially if Nature has chosen a value of $r \gtrsim 0.005$, which is close to the sensitivity limit found in this study. Other complementary information concerning $\beta_s$ and/or $\alpha_s$, as those coming from future planned galaxy surveys \cite{Basse:2014qqa} (for instance, the SPHEREX project~\cite{Dore:2014cca}), could significantly improve the sensitivities forecasted here. 

{\em Acknowledgments. ---}
The authors would like to thank L.~Verde for useful comments on the manuscript. OM is supported by PROMETEO II/2014/050, by the Spanish Grant FPA2011--29678 of the MINECO and by PITN-GA-2011-289442-INVISIBLES. LB and HR acknowledge financial support from PROMETEO II/2014/050. ME is supported by Spanish Grant FPU13/03111 of MECD. LB, ME and HR acknowledge the warm hospitality of the HECAP section of the ICTP, where part of this work was done.

\section{Appendix}

\subsection{Consistency of the Fisher Method}
\label{sec:ap1} 
We test the validity of our method by computing our Fisher matrix forecast for the complete \emph{Planck} mission and comparing our results to those obtained by \emph{Planck} measurements. For that purpose, we shall use the $100$, $143$ and $217$ GHz channels of \emph{Planck} with its accounted foreground removal as shown in Table \ref{tab:Foregroundstab}, and following the specifications detailed in Table~\ref{Table:Experiments} of the Appendix B. From the results depicted in Table \ref{Table:ForecastComparison}, notice that there is an excellent agreement between the forecasted parameter errors and the errors quoted by the \emph{Planck} collaboration, with the differences always below the $20\%$ level. In addition, we have verified that the correlations between the cosmological parameters are well accounted for. 
\subsection{CMB Mission specifications}
Table \ref{Table:Experiments} shows the values used for the sensitivity of \emph{Planck} and COrE missions, as in Ref.~\cite{Bouchet:2011ck}.

\label{sec:ap2} 
\begin{table}[t] 
\begin{center}
\begin{tabular}{lcccc}
\toprule
Mission & Channel & FWHM  & $\Delta T $  & $\Delta P$ \\ 
& (GHz) & (arcmin) &   ($\mu K_{\rm{CMB}} \cdot$arcmin) & ($\mu K_{\rm{CMB}} \cdot$arcmin) \\
\hline 
& 30  & 32.7 & 203.2 & 287.4 \\
& 44 & 27.9 & 239.6 & 338.9  \\
& 70 & 13.0 & 221.2 & 298.7 \\
& 100 & 9.9 & 31.3 & 44.2 \\
\emph{Planck} & 143 & 7.2 & 20.1 & 33.3 \\
& 217 & 4.9 & 28.5 & 49.4 \\
& 353 & 4.7 & 107.0 & 185.3 \\
& 535 &	4.7 & 1100 & - \\
& 857 &	4.4 & 8300 & -  \\
\hline
& 45 & 23.3 & 5.25 & 9.07 \\ 
& 75 & 14.0 & 2.73 & 4.72\\ 
& 105 & 10.0 & 2.68 & 4.63\\ 
& 135 & 7.8 & 2.63 & 4.55\\ 
& 165 & 6.4 & 2.67 & 4.61\\ 
& 195 & 5.4 & 2.63 & 4.54\\ 
& 225 & 4.7 & 2.64 & 4.57\\ 
CORE & 255 & 4.1 & 6.08 & 10.5\\ 
& 285 & 3.7 & 10.1 & 17.4\\ 
& 315 & 3.3 & 26.9 & 46.6\\ 
& 375 & 2.8 & 68.6 & 119\\ 
& 435 & 2.4 & 149 & 258\\ 
& 555 & 1.9 & 227 & 626\\ 
& 675 & 1.6 & 1320 & 3640\\ 
& 795 & 1.3 & 8070 & 22200 \\
\hline \hline 
\end{tabular}
\caption{Specifications of the \emph{Planck} and COrE experiments, from \cite{Bouchet:2011ck}.}
\label{Table:Experiments}
\end{center}
\end{table}

\end{document}